\pdfoutput=1
\documentclass[
  reprint,
  superscriptaddress,
  amsmath,amssymb,
  aps,
  showkeys,
]{revtex4-2}

\usepackage{graphicx}
\usepackage{dcolumn}
\usepackage{siunitx}
\usepackage[T1]{fontenc}

\usepackage[version=4]{mhchem}

\usepackage{color,soul}

\DeclareSIUnit\Molar{\textsc{m}} 
\DeclareSIUnit{\wtpercent}{wt.\%}
\DeclareSIUnit\cal{cal}

\begin{document}
\title{Powder-in-tube confinement as a design principle for fatigue-resistant caloric materials}

\author{T. Niehoff}%
\email{t.niehoff@hzdr.de}
\affiliation{Dresden High Magnetic Field Laboratory (HLD-EMFL) and Würzburg-Dresden Cluster of Excellence ctd.qmat, Helmholtz-Zentrum Dresden-Rossendorf, 01328 Dresden, Germany}

\author{L.~Beyer}%
\affiliation{Leibniz Institute for Solid State and Materials Research Dresden, 01069 Dresden, Germany}
\affiliation{Institute of Materials Science, Technische Universität Bergakademie Freiberg, 09599 Freiberg, Germany}

\author{J.~Puy}%
\affiliation{Institute of Materials Science, Technical University of Darmstadt, 64287 Darmstadt, Germany}

\author{F.~Scheibel}%
\affiliation{Institute of Materials Science, Technical University of Darmstadt, 64287 Darmstadt, Germany}
 
\author{J.~Freudenberger}%
\affiliation{Leibniz Institute for Solid State and Materials Research Dresden, 01069 Dresden, Germany}
\affiliation{Institute of Materials Science, Technische Universität Bergakademie Freiberg, 09599 Freiberg, Germany}
 
\author{O.~Gutfleisch}%
 \affiliation{Institute of Materials Science, Technical University of Darmstadt, 64287 Darmstadt, Germany}

\author{J.~Wosnitza}%
\affiliation{Dresden High Magnetic Field Laboratory (HLD-EMFL) and Würzburg-Dresden Cluster of Excellence ctd.qmat, Helmholtz-Zentrum Dresden-Rossendorf, 01328 Dresden, Germany} 
\affiliation{Institut für Festkörper- und Materialphysik, Technische Universität Dresden, 01069 Dresden, Germany}

\author{T.~Gottschall}
\affiliation{Dresden High Magnetic Field Laboratory (HLD-EMFL) and Würzburg-Dresden Cluster of Excellence ctd.qmat, Helmholtz-Zentrum Dresden-Rossendorf, 01328 Dresden, Germany}

\date{\today}

\begin{abstract}

Caloric solid-state refrigeration offers an energy-efficient alternative to vapor-compression cooling, but many of the most promising materials, including the magneto- and elastocaloric Heusler alloys, are intrinsically brittle and fail under the cyclic loading required in operation. Here, we introduce mechanical confinement via powder-in-tube (PIT) processing as a general strategy to overcome this limitation. By embedding a granular Heusler alloy core of \SI{87}{\percent} packing density within a ductile steel sheath (\SI{2.8}{\milli\meter} outer diameter, \SI{0.3}{\milli\meter} wall thickness), the composite confines the brittle powder and enables effective load transfer between particles. While the core remains inherently brittle, the surrounding sheath prevents catastrophic fragmentation and turns brittleness from a disqualifying property into a manageable one. We demonstrate that this approach enhances mechanical durability by several orders of magnitude, with the composites sustaining 100,000 load cycles at \SI{250}{\mega\pascal} without structural failure and tolerating overload stresses up to \SI{700}{\mega\pascal}. Confinement also alters the transformation itself, and to resolve this behavior in these mechanically complex systems, we developed a simultaneous in-situ measurement technique combining strain and AC magnetic susceptibility under load. It reveals that the internal stress field in the core is strongly heterogeneous. As a consequence, the macroscopic strain decouples from the transforming phase fraction, and the martensitic transition cannot be tracked by mechanical data alone. Mechanical confinement thus emerges as a design principle for fatigue-resistant caloric materials, and the combined in-situ method as a characterization tool for composite, porous, and other mechanically heterogeneous caloric systems.

\end{abstract}

\newpage


\maketitle

\section{\label{sec:Introduction}Introduction}

Solid-state refrigeration based on caloric effects has emerged as a promising alternative to conventional vapor-compression technologies due to its high energy efficiency and environmental compatibility \cite{MEVADA202486, https://doi.org/10.1002/ente.201800137, Kitanovski2020, https://doi.org/10.1002/aenm.202400369}. Among the candidate materials, magnetic shape-memory alloys of the Heusler family are particularly attractive because they exhibit large magnetocaloric and elastocaloric effects associated with a first-order magnetostructural phase transition between a ferromagnetic austenite and a weakly magnetic martensite \cite{PFEUFFER2021117390, CHEN2021102706, QIAN20161, 10.3389/fmats.2022.812984}.

Despite their outstanding functional properties, the practical application of these materials is severely limited by their intrinsic brittleness and by pronounced functional fatigue, especially in polycrystalline bulk material \cite{ma16165725}. Under the cyclic mechanical loading required for caloric operation, Heusler alloys are highly susceptible to intergranular fracture and subsequent catastrophic failure \cite{SCHEIBEL2023101783}. This combined loss of structural and functional integrity under cyclic loading represents a fundamental barrier to the implementation of Heusler alloys in cooling devices.

Previous efforts to address this limitation have primarily focused on improving the intrinsic mechanical properties of bulk materials through alloy design, microstructural optimization, or advanced processing routes \cite{CHEN2021102706}. Approaches such as vacuum casting \cite{ZHANG2024119595}, sintering \cite{VILLA20245582, WU2025416611}, or directional solidification have demonstrated enhanced mechanical stability, in part through advantageous crystallographic texturing that improves resistance to compressive stress \cite{PFEUFFER2021117157, GUAN2023115353, ZHANG2023115725, ma17194756}. Furthermore, strategies involving doping or grain-boundary engineering have been explored to toughen the microstructure and arrest crack propagation \cite{IMRAN2021107255, DESOUZASILVA20235167, QIAN2024113162, PFEUFFER2021117390}. However, these strategies often involve complex fabrication routes and typically lead to trade-offs between fatigue-resistant and caloric performance. Moreover, bulk materials remain inherently prone to fatigue failure under repeated high-stress loading. A further approach reported in the literature is to optimize the crystal-lattice compatibility between the austenite and martensite phases, thereby minimizing the thermal hysteresis \cite{Cui2006, ZHANG20094332}, which in turn improves the functional fatigue resistance \cite{MEVADA202486, PhysRevApplied.17.054032}. Here again, this improvement is often achieved at the expense of caloric performance. Moreover, multicaloric cooling cycles that deliberately exploit thermal hysteresis \cite{Gottschall2018, 10.1063/5.0238689} require a large hysteresis, which renders this optimization strategy unsuitable for such applications.

In contrast, an alternative strategy is to stabilize the material externally rather than intrinsically. The powder-in-tube (PIT) method follows this route by encapsulating the functional material within a ductile metallic sheath. While the technique itself is well established, its previous use has been confined to applications, in which the sheath protects or stabilizes the core against handling, drawing, or environmental degradation, rather than against functional cyclic loading. In superconducting-wire fabrication, PIT mechanically stabilizes the otherwise brittle superconducting compounds during processing and winding \cite{NAST20021241, KOVAC201940, Yao_2019, BALAMURUGAN20041184, Ma_2012}. In the field of caloric materials, it has been used to produce La(Fe,Co,Si)$_{13}$ composite wires \cite{FUNK2018223, KRAUTZ2020108832} and to protect gadolinium cores from corrosion upon contact with a heat-transfer fluid \cite{BEYER2021168115}.

Here, we extend the PIT concept to a fundamentally different loading regime. We mechanically confine a brittle Heusler alloy under uniaxial cyclic compression, as required for elastocaloric operation. In this configuration the ductile sheath imposes a confining stress state on the granular core, preserving the integrity of the core even as individual particles crack, maintaining load transfer across particle contacts, and preventing the catastrophic failure that limits bulk elastocaloric materials. At the same time, the concept enables the use of powder-based materials that would otherwise be mechanically unstable in bulk form.

While PIT processing thus offers a promising route toward mechanically fatigue-resistant elastocaloric materials, it also introduces new challenges for the characterization of the phase transformation. Unlike in bulk materials, the macroscopic strain response of the composite is governed by the mechanical behavior of the sheath and by the heterogeneous internal stress distribution within the core. Conventional stress-strain measurements are, therefore, insufficient to determine the transformation behavior and phase fraction directly.

To address this challenge, we developed a simultaneous in-situ measurement technique that combines mechanical-strain application and AC magnetic-susceptibility detection under uniaxial load. This approach allows direct access to the functional transformation independent of the macroscopic mechanical response.

In this work, we demonstrate that mechanical confinement via PIT processing provides a robust pathway to overcome the intrinsic brittleness of Heusler alloys. The composite architecture significantly enhances mechanical durability under cyclic uniaxial compression and, at the same time, fundamentally modifies the transformation behavior through the heterogeneous internal stress fields it introduces. To resolve this modified transformation independently of the composite's macroscopic mechanical response, we employ a simultaneous in-situ technique combining uniaxial loading with AC magnetic-susceptibility detection. Together, these results establish composite confinement as a design principle for mechanically robust caloric materials under uniaxial load and introduce the in-situ methodology required to characterize the transformation behavior it induces.

\section{\label{sec:Experimental Section}Experimental}

\begin{figure*}[t]
  \centering
  \includegraphics{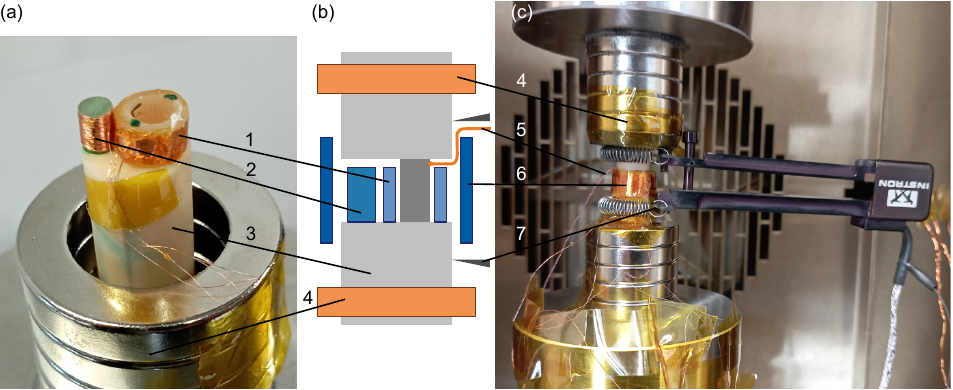}
  \caption{(a) Detailed view of the lower part of the simultaneous susceptibility and stress-strain measurement setup. (b) Schematic sketch and (c) photograph of the complete assembly within the climate chamber. The numbered components are: (1) Measurement coil (50 turns, \SI{6}{\milli\meter} diameter). (2) Compensation coil (\SI{3}{\milli\meter} diameter, 110 turns, in the opposite direction and connected in series). (3) Ceramic pistons (\SI{10}{\milli\meter} diameter) between which the coils and the sample are clamped. (4) Permanent magnets (generating a \SI{140}{\milli\tesla} field along the $z$ direction at the sample). (5) Thermocouple (measures sample temperature). (6) Outer solenoid (inner diameter \SI{10.5}{\milli\meter}, outer diameter \SI{12}{\milli\meter}, 200 turns, generates the AC field at \SI{1.0487}{\kHz}). (7) Extensometer (measures length change/displacement).}
  \label{fig:Darmstadt_Aufbaue}
\end{figure*}

\subsection{Sample preparation}

We synthesized a polycrystalline Heusler alloy with the nominal composition Ni$_{45}$Co$_{2.5}$Fe$_{2.5}$Mn$_{38.5}$Sn$_{11.5}$ using arc melting under an argon atmosphere. We utilized high-purity elements as starting materials: Ni (\SI{99.995}{\percent}), Co (\SI{99.95}{\percent}), and Sn (\SI{99.998}{\percent}), supplied by Kurt J. Lesker, as well as Fe (\SI{99.97}{\percent}) and Mn (\SI{99.99}{\percent}), supplied by chemPUR.

Prior to the alloying process, we purified the manganese pieces by pre-melting them to remove volatile surface impurities and oxides. To compensate for Mn evaporation losses typical during arc melting, we added an excess of \SI{3}{\wtpercent} Mn to the initial stoichiometric weight. We prepared an ingot with a mass of approximately \SI{5}{\gram}. To ensure chemical homogeneity, we remelted the sample four times, flipping the ingot between each melting step.

Subsequently, the as-cast ingot was annealed in an argon atmosphere at \SI{950}{\degreeCelsius} for \SI{24}{\hour}, a temperature slightly below the melting temperature of the alloy, followed by water quenching to room temperature. The homogenized bulk sample served as the reference material for this study and as the precursor for the PIT processing.

\subsection{Fabrication of powder-in-tube (PIT) samples}

To mechanically stabilize the samples for elasto- and magnetocaloric measurements, we fabricated composite specimens using the PIT method.

The fabrication followed a multi-step protocol:

\begin{enumerate}
    \item Arc melting: We produced the initial sample as described above.
    \item Pulverization: We mechanically crushed the homogenized bulk ingots using a mortar and pestle. The resulting powder was then sieved to obtain particles smaller than \SI{100}{\micro\meter}.
    \item Filling: We used a seamless austenitic stainless-steel (316L) tube with an initial outer diameter of \SI{6}{\milli\meter} and an inner diameter of \SI{4}{\milli\meter}. Prior to filling, one end of the tube was sealed by swaging the first \SI{2}{\centi\meter} over a brass plug. The pre-alloyed powder was then filled into the tube to a height of \SI{50}{\milli\meter}. Finally, the opposite end was secured with a slotted plug via rotary swaging; the slot was necessary to allow any entrapped gas to escape during subsequent processing.
    
    \item Swaging: The filled tube was cold rotary swaged on a UR-2-4 machine (HMP) in six consecutive passes, each with an area reduction of approximately \SI{20}{\percent}, following the procedure reported by Funk et al.~\cite{FUNK2018223}. This process reduced the outer diameter to approximately \SI{3.1}{\milli\meter}, resulting in a steel-wall thickness of \SI{0.6}{\milli\meter} (degree of deformation $\varphi_A = 1.32$).
\end{enumerate}

This cold work compacted the powder into a cohesive, stable core. X-ray computed tomography (XCT) yielded a packing density of approximately \SI{87}{\percent} for the functional Heusler core.

To optimize the magnetocaloric and elastocaloric efficiencies, it is crucial to minimize the thermal mass of the inactive sheath material. Therefore, we turned down the steel sheath on a lathe, removing \SI{0.3}{\milli\meter} of the nominal \SI{0.6}{\milli\meter} wall thickness. Following the mechanical processing, we subjected the samples to final annealing treatments to relieve deformation-induced stresses and restore the magnetic properties of the Heusler core. Since optimizing this restoration is a key aspect of the present work, the specific annealing protocols and their impact on the sample properties are discussed in the Results section.

The machined samples had a final outer diameter of approximately \SI{2.8}{\milli\meter} and a height of \SI{5.5}{\milli\meter}. We ground the sample ends flat and parallel to ensure a uniform load distribution during uniaxial compression. The total mass of a final PIT sample, including the steel sheath, was approximately \SI{260}{\milli\gram}.

\subsection{Characterization procedures}
We used XCT to determine the filling factor of the Heusler alloy core and to visualize the internal structure and crack formation after several loading cycles on a macroscopic scale. The XCT scans were performed with a Phoenix Nanotom m 180 (General Electric). To ensure comparability, all samples were scanned equally using the following parameters: we applied a voltage of \SI{140}{\kilo\volt} and a current of \SI{120}{\micro\ampere} with an exposure time of \SI{1000}{\milli\second} for each of the 1080 images, filtered by a \SI{0.3}{\milli\meter} Cu sheet. The sample-to-detector configuration resulted in a voxel size of \SI{3.6}{\micro\meter}.

We reconstructed the resulting two-dimensional images into a three-dimensional
volume using Phoenix datosjx2 software (Waygate Technologies). To assess the filling factor and depict the progression of the internal structure after the compression testing, VGStudio MAX (Volume Graphics) was used.

We examined the microstructure of the samples using a Hitachi S-4800 scanning electron microscope (SEM). For imaging, samples were sectioned parallel and perpendicular to the tube axis, then ground and polished. The magnetization was measured using a vibrating sample magnetometer (Quantum Design PPMS).

\subsection{Simultaneous susceptibility and stress-strain measurement setup}

To evaluate the elastocaloric properties and the degree of transformation, we performed uniaxial compression tests using an Instron 5967 equipped with an environmental chamber. This setup allowed us to apply a defined load while controlling the ambient temperature between \SI{180}{} and \SI{350}{\kelvin}. The applied stress was calculated from the applied force and the initial cross-sectional area of the sample.

Preliminary investigations revealed that the stress-strain behavior of the PIT composite samples deviated significantly from that of bulk materials, preventing a clear identification of the onset of the structural transition based on mechanical data alone. To resolve the phase transition over the entire volume of the composite, we developed a novel experimental setup capable of measuring the magnetic susceptibility simultaneously with mechanical loading.

Figure 1(a) shows a photograph of the lower section of the custom-built setup, while Fig. 1(c) shows a picture of the setup installed within the environmental chamber. Figure 1(b) shows a schematic sketch of the assembly.

The measurement principle and components are described as follows: The sample (dark grey) is positioned inside a measurement coil (1) having an inner diameter of \SI{6}{\milli\meter} and 50 turns. To suppress the background signal, a compensation coil (2) is connected to coil (1) in opposite direction. Due to spatial constraints, this compensation coil was designed with a reduced diameter of \SI{3}{\milli\meter} and 110 turns. Although approximately 200 turns would be required for full compensation, the available winding length only accommodates 110 turns. The chosen configuration nevertheless provides a sufficient signal-to-noise ratio.

The coils and the sample are clamped between two ceramic pistons (3) (\SI{10}{\milli\meter} diameter), which transmit the uniaxial load to the sample. To ensure a defined magnetic state, permanent magnets (4) are arranged to generate a static bias field of \SI{140}{\milli\tesla} along the $z$ direction (loading axis), aligning the magnetic domains within the polycrystalline Heusler core.

The sample temperature is monitored in-situ by a thermocouple (5) attached directly to the sheath. The AC magnetic susceptibility is probed using an outer excitation solenoid (6) (inner diameter \SI{10.5}{\milli\meter}, outer diameter \SI{12}{\milli\meter}, 200 turns), which generates an alternating magnetic field at a frequency of \SI{1.0487}{\kHz}.

Finally, the strain is measured using an clip-on extensometer (7). The extensometer knife-edges are attached to the upper and lower ceramic piston via spring clips. By recording the displacement of the pistons relative to the initial sample length, the total strain of the composite can be determined alongside the magnetic response.

\section{\label{sec:Diskussion}Results and Discussion}

\subsection{Heat treatment}

\begin{figure}[t]
  \centering
  \includegraphics{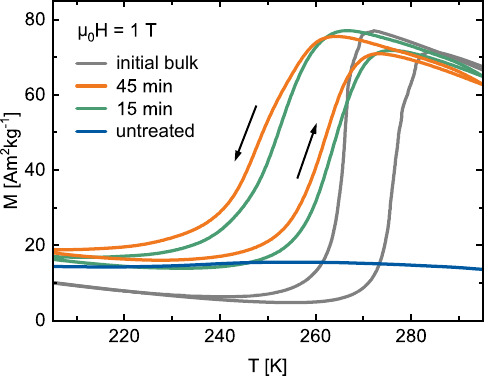}
  \caption{Temperature-dependent magnetization measured by use of a vibrating-sample magnetometer at \SI{1}{\tesla} for PIT samples after various heat-treatment durations (untreated, 15, and \SI{45}{\minute}) at \SI{950}{\degreeCelsius}. The curve for the bulk reference sample (annealed for  \SI{24}{\hour}) is included for comparison.}
  \label{fig:heat_treatment_M(T)}
\end{figure}

\begin{figure*}[t]
  \centering
  \includegraphics{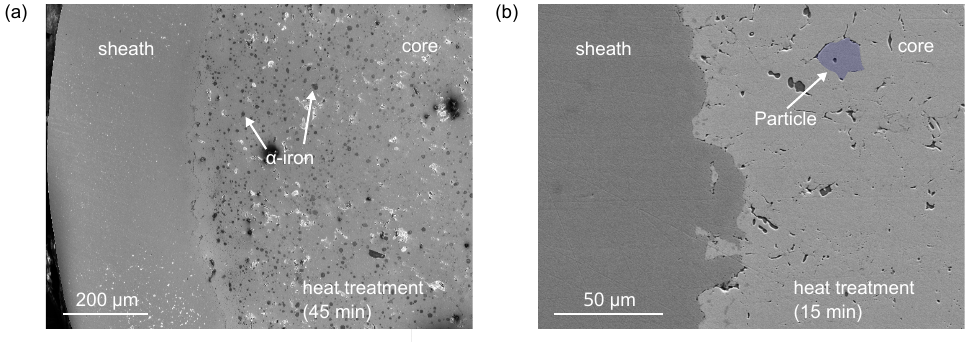}
    \caption{(a) Transverse SEM cross-section of a PIT sample after prolonged heat treatment (45 min.), where secondary phases are visible as dark spots near the steel sheath. (b) Longitudinal SEM cross-section of a PIT sample. Individual powder particles from the milling process remain partially distinguishable, one is highlighted in blue as an example.}
  \label{fig:heat_treatment}
\end{figure*}

Internal stress and defects introduced by the mechanical pulverization of the Heusler alloy and the PIT process lead to a loss of the desired functional magnetic properties, most notably the suppression of the austenitic phase transition \cite{CAVAZZINI2021159747}. To restore the crystalline structure and functional properties, we subjected the PIT samples to a recovery heat treatment at \SI{950}{\degreeCelsius}, similar to the treatment applied to the bulk material.

Figure~\ref{fig:heat_treatment_M(T)} shows the magnetization measured at \SI{1}{\tesla} as a function of temperature for the untreated and heat treated PIT samples. The M(T) curve of the initial bulk material is included for reference. Without any heat treatment (untreated), the structural transition is completely suppressed. This can be attributed to the high residual lattice strain and high defect density induced during the milling process, which effectively stabilize the martensite phase \cite{SANCHEZALARCOS2016983, SANCHEZALARCOS2021159536}. However, after only a short annealing period of \SI{15}{\minute}, the PIT sample already reaches nearly the same maximum magnetization across the magnetostructural transition as the bulk reference.

Prolonged heat treatment does not further improve the saturation magnetization. On the contrary, treatment beyond \SI{15}{\minute}, as shown by the 45-min curve, leads to an undesirable broadening of the thermal hysteresis towards lower temperatures. This behavior is likely related to the diffusion of iron from the surrounding steel sheath into the Heusler core. An increased Fe content is known to shift the phase transition to lower temperatures \cite{DELTELL20211091, Younsi2024}. Consequently, the region of the PIT core adjacent to the sheath exhibits a transformation at reduced temperatures, causing the overall transition to broaden.

Furthermore, the enhanced iron content promotes the formation of iron-rich secondary phases. We observed a high density of these secondary phases near the sheath in the SEM cross-section image of the 45-min. heat-treated PIT sample [Fig.~\ref{fig:heat_treatment}(a)]. These secondary phases appear as small, dark spots within the core material, with their density decreasing deeper into the core.

A second crucial aspect of the heat treatment concerns the magnetic background signal originating from the steel sheath. This undesired signal, visible in Fig.~\ref{fig:heat_treatment_M(T)} as an offset of approximately \SI{10}{A m^2 kg^{-1}} in the PIT curves, is induced by the cold rotary swaging and subsequent martensite formation within the sheath itself. Eliminating this background signal entirely typically requires annealing the steel at \SI{1050}{\degreeCelsius} for 10 minutes \cite{FUNK2018223}. However, this is not feasible for the PIT composites, as the Heusler core material begins to melt above \SI{1000}{\degreeCelsius}, which would destroy the functional core.

Despite originating from the same bulk material, the phase transition of the 15-min-annealed PIT sample is shifted by approximately \SI{15}{\kelvin} to lower temperatures compared to the bulk reference (Fig.~\ref{fig:heat_treatment_M(T)}). Since no significant compositional changes are expected during milling or swaging, the observed behavior is attributed to lattice distortions induced by deformation and confinement. These distortions generate a stress field that energetically favors the martensitic state.

A longitudinal SEM cross-section of the 15-min-annealed PIT sample [Fig.~\ref{fig:heat_treatment}(b)] confirms that swaging compacts the milled powder into a dense, cohesive core, with individual powder particles still partially distinguishable. The core bonds to the steel sheath along most of the interface, with only a few small gaps remaining. This close contact is a prerequisite for efficient load transfer between particles and for the thermal coupling needed in a cooling cycle.

Based on these results, we selected a heat-treatment duration of 15 minutes at \SI{950}{\degreeCelsius} for all subsequent investigations, as this protocol effectively restores the magnetic transition while minimizing detrimental secondary phase formation and transition broadening.

\subsection{Cyclic loading and maximum stress}

\begin{figure*}[t]
  \centering
  \includegraphics{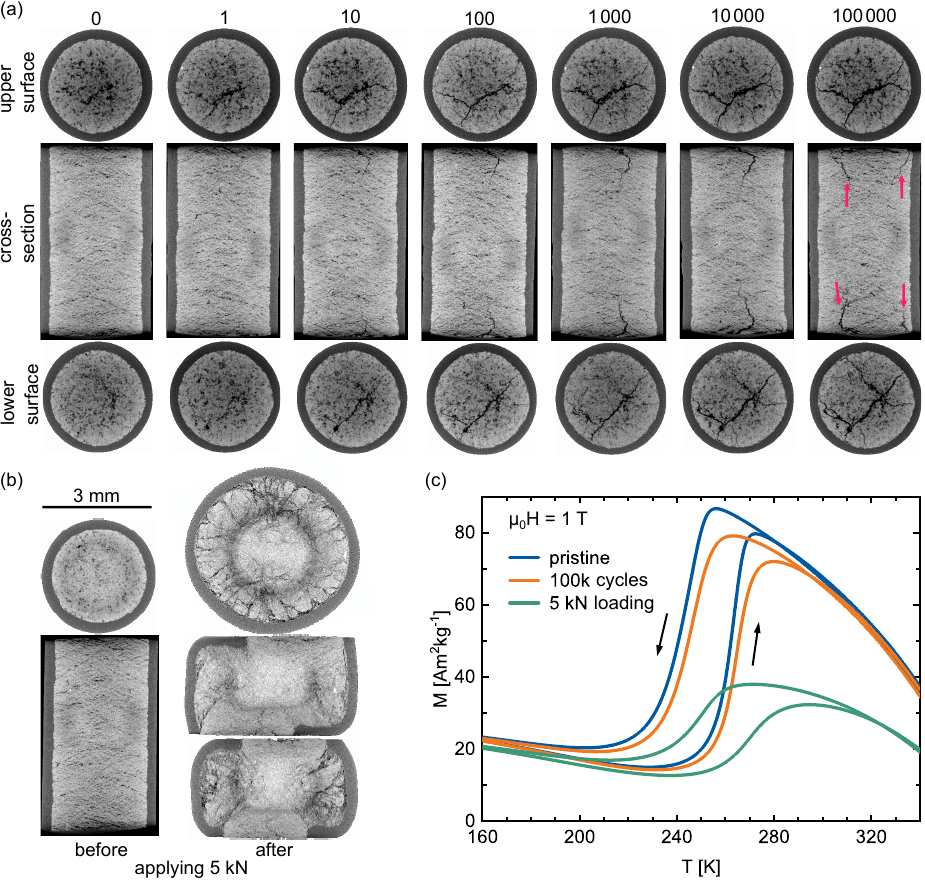}
  \caption{(a) X-ray computed tomography (XCT) cross-sections of a powder-in-tube (PIT) sample demonstrating core integrity under cyclic compressive load (\SI{250}{\mega\pascal} at \SI{250}{\kelvin}). For each state, the top and bottom panels show horizontal cross-sections taken at the upper and lower end of the sample, respectively, while the central panel shows a vertical cross-section through the sample centre. Scans are shown before loading and after 1, 10, 100, 1000, 10\,000, and 100\,000 cycles. Minor fracturing is visible only at the core ends (pink arrows). (b) XCT scans of a PIT sample before and after application of the maximum load (\SI{5000}{\newton}, corresponding to approximately \SI{700}{\mega\pascal}), revealing pronounced plastic deformation without structural fracture. The pristine state is shown as one central vertical and one horizontal cross-section; the post-load state is shown as one horizontal cross-section together with two orthogonal vertical cross-sections through the sample centre. (the sample had a pre-history of 30,000 cycles at \SI{250}{\mega\pascal}.) (c) Temperature-dependent magnetization of two PIT samples. For the first sample, the pristine state (0 cycles) is compared with the state after 100,000 load cycles at \SI{250}{\mega\pascal}; the second sample is shown after the maximum-load test.}
  \label{fig:PIT_Druck}
\end{figure*}

The primary motivation for utilizing the PIT technique is to overcome the inherent brittleness of Heusler alloys and other elastocaloric materials by mechanically confining the fractured core, thereby achieving fatigue-resistant composites suitable for practical applications. We, therefore, investigated the mechanical durability of the PIT samples under cyclic and maximal compressive loading.

To assess the cyclic mechanical reliability, we subjected a PIT sample to 100,000 load cycles under a uniaxial stress of \SI{250}{\mega\pascal} on an Instron 8502 servo-hydraulic testing machine at a constant temperature of \qty{250}{\kelvin}. To reduce experimental time, we did not execute the full phase-transformation protocol (i.e., cooling the sample into the martensitic phase, as required for a complete elastocaloric cycle) between each load cycle.

To nondestructively monitor crack formation and potential core delamination, we performed XCT scans before loading and after 1, 10, 100, 1000, 10\,000, and 100\,000 cycles. The results are presented in Fig.~\ref{fig:PIT_Druck}(a).

If the core detaches from the sheath due to internal cracking, the essential thermal coupling between core and sheath, which is required for the cooling cycle, would be compromised. However, the XCT images show that crack formation is confined to minor fracturing at the upper and lower ends of the core only [Fig.~\ref{fig:PIT_Druck}(a)]. These localized cracks do not compromise the overall structural integrity or the feasibility for practical application.

Furthermore, to evaluate whether cyclic loading alters the functional magnetic properties, Fig.~\ref{fig:PIT_Druck}(c) compares the temperature-dependent magnetization before the first compression and after 100,000 load cycles. We observed only minor changes, indicating that the functional properties are not significantly affected by prolonged cyclic stress at this level.

To study the maximum load-bearing capacity, we compressed a PIT sample at \SI{250}{\kelvin} with an engineering stress of approximately \SI{700}{\mega\pascal}. This specific sample had already undergone 30\,000 cycles at \SI{250}{\mega\pascal} prior to the maximum load test. Typical Heusler bulk samples fracture at significantly lower stresses. In contrast, the PIT sample does not fracture but instead undergoes substantial plastic deformation [Fig.~\ref{fig:PIT_Druck}(b)]. The final diameter of the sample increased by approximately \SI{50}{\percent}, while the height is roughly halved compared to the original geometry.

Under overloading the steel sheath yields plastically instead of fracturing, so that the powder core remains confined and the sample deforms rather than disintegrating into loose powder, which would render it unusable in a refrigeration circuit.

However, this maximal mechanical damage impacts the functional properties. Figure~\ref{fig:PIT_Druck}(c) shows that the magnetization at the magnetostructural transition is reduced by more than \SI{50}{\percent} following the maximal load test, compromising the functionality. The magnetic properties are expected to be at least partially recoverable by a subsequent short annealing treatment.

\subsection{Simultaneous measurement}

For the elastocaloric effect, it is necessary to know the strain required to fully transform the material (here, austenite to martensite). For the following measurements, we used our purpose-built setup.

\subsubsection{Isostress measurement}

\begin{figure}[t]
  \centering
  \includegraphics{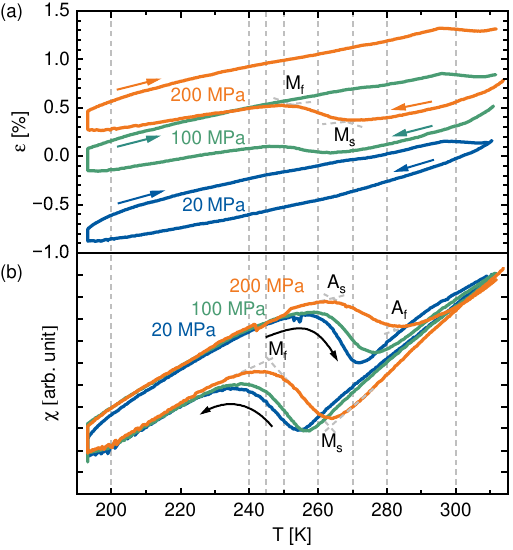}
  \caption{(a) Temperature-dependent strain ($\varepsilon$) and (b) AC magnetic susceptibility ($\chi$) of a PIT sample measured simultaneously under constant uniaxial stresses of \SI{20}{\mega\pascal}, \SI{100}{\mega\pascal}, and \SI{200}{\mega\pascal}. The characteristic transition temperatures for martensite start ($M_s$), martensite finish ($M_f$), austenite start ($A_s$), and austenite finish ($A_f$) are indicated. Grey dashed lines mark the temperatures, at which the isothermal measurements shown in Fig.~\ref{fig:PIT_Stress_strain(F)} were performed.}
  \label{fig:PIT_Stress_strain(T)}
\end{figure}

Figure~\ref{fig:PIT_Stress_strain(T)}(a) displays the temperature-dependence of the strain ($\varepsilon$) under various constant uniaxial stresses: \SI{20}{\mega\pascal}, \SI{100}{\mega\pascal}, and \SI{200}{\mega\pascal}. Upon cooling, the transition from austenite to martensite ($M_s$) is marked by an increase in strain. While this effect is very weak at \SI{20}{\mega\pascal}, the magnitude of the strain change increases with increasing pressure, and the transition begins at higher temperatures. This shift and broadening align with the expected behavior, where pressure favors the martensitic phase, which possesses a smaller volume.

The subsequent heating cycle shows no clear reverse transformation (martensite to austenite) in the strain signal across the measured temperature range (only a very faint indication is visible at \SI{20}{\mega\pascal}).

To investigate the cause of this apparent absence of the reverse transformation in the strain signal, we simultaneously measured the AC susceptibility ($\chi$), as shown in Fig.~\ref{fig:PIT_Stress_strain(T)}(b). Here, the structural transition is clearly visible upon cooling for all applied stresses, even at \SI{20}{\mega\pascal}.

Notably, we observe a shift in the characteristic transition temperatures ($M_s$ and $M_f$) when comparing the simultaneous measurements. For instance, at \SI{200}{\mega\pascal}, $M_s$ derived from the strain signal is about \SI{268}{\kelvin}, while the susceptibility measurement reveals this transition at about \SI{5}{\kelvin} lower temperature. The shift at $M_f$ is even larger, approximately \SI{8}{\kelvin} lower than the \SI{252}{\kelvin} derived from the strain data.

Comparing the susceptibility results at \SI{20}{\mega\pascal} ($M_s$ $\approx$\,\SI{254}{\kelvin}, $M_f$ $\approx$\,\SI{237}{\kelvin}) with magnetization data reveals further shifts, confirming a dependence on the measurement technique and internal state. Given the long measurement duration ($\approx$\,\SI{3.5}{\hour}, corresponding to $\approx$\,\SI{1}{\kelvin\per\minute}), dynamic effects are unlikely to be the primary cause of these discrepancies.

We propose that the substantial shift in the observed transition temperatures is rooted in the heterogeneous stress distribution within the composite core, caused by the PIT confinement and porosity. This non-uniform stress field leads to a complex strain response that prevents a direct quantitative correlation between the macroscopic strain and the transformed phase fraction. This assumption is supported by simultaneous measurements on Heusler bulk samples \cite{BECKMANN2025120460}, which showed no such deviations between magnetization and strain-derived transition temperatures. This strongly suggests that the observed discrepancy is primarily induced by the internal stresses inherent to the PIT composite.

Upon heating, the susceptibility signal clearly reveals the reverse transformation (austenite start, $A_s$, and finish, $A_f$), which also shifts symmetrically to higher temperatures under pressure. The offset at the temperature reversal point ($\approx$\,\SI{193}{\kelvin}) in both signals is an artifact caused by the thermal inertia of the environmental chamber during the stabilization period. Likewise, the feature appearing in the strain signal at around \SI{295}{\kelvin} upon heating is of instrumental origin and is not related to the martensitic transformation.

\subsubsection{Isothermal measurement}

\begin{figure*}[t]
  \centering
  \includegraphics{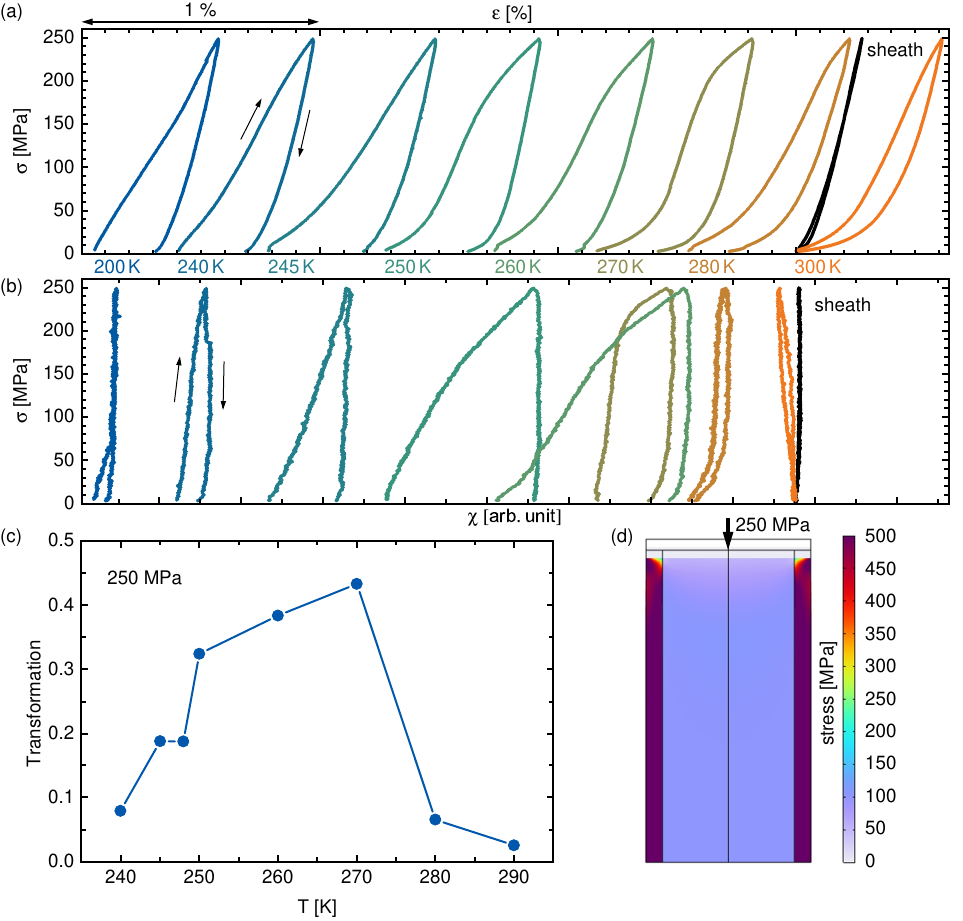}
  \caption{Simultaneous isothermal measurements of a PIT sample at the temperatures indicated by dashed lines in Fig.~\ref{fig:PIT_Stress_strain(T)}. (a) Selected stress-strain curves and (b) AC magnetic susceptibility ($\chi$) as a function of uniaxial stress up to \SI{250}{\mega\pascal}. The response of the empty steel sheath at \SI{300}{\kelvin} is shown for reference. (c) Estimated martensitic transformation fraction derived from the isothermal and isostress measurements. (d) Finite-element simulation of a PIT sample under \SI{250}{\mega\pascal} uniaxial compression, performed with COMSOL Multiphysics, showing the deformed cross-section with displacements exaggerated ten times for visualization. The original geometry is indicated by outlines and the upper rectangle represents the loading piston. Colors indicate the local axial stress distribution. The simulation assumes linear elastic behavior with Young's moduli of \SI{200}{\giga\pascal} for the steel sheath and \SI{60}{\giga\pascal} for the porous Heusler core, without accounting for the magnetostructural transition. The total sample radius is approximately \SI{1.5}{\milli\meter}. The steel sheath, initially \SI{0.6}{\milli\meter} thick, was turned down by about \SI{0.3}{\milli\meter}, leaving a nominal wall thickness of \SI{0.3}{\milli\meter}.}
  \label{fig:PIT_Stress_strain(F)}
\end{figure*}

We also measured the strain ($\varepsilon$) and susceptibility ($\chi$) under isothermal load cycles up to \SI{250}{\mega\pascal} at the temperatures. The corresponding temperatures are indicated by dashed lines in Fig.~\ref{fig:PIT_Stress_strain(T)}. Prior to the load cycles, we followed a thermal protocol, in which the sample was first heated to \SI{300}{\kelvin}, well above the transition, to erase any stress-induced martensite and then cooled to the measurement temperature, resulting in a predominantly austenitic starting state.

Figure~\ref{fig:PIT_Stress_strain(F)}(a) shows the isothermal stress-strain data and Fig.~\ref{fig:PIT_Stress_strain(F)}(b) the simultaneously measured susceptibility-stress curves.

Based on literature data for Heusler bulk samples \cite{PFEUFFER2021117157}, one expects that at temperatures near $A_s$, the strain and, therefore, the susceptibility should increase sharply upon compression until the sample is fully converted to martensite, followed by a transition into a linear elastic region. While the PIT sample exhibits a broadened and less pronounced strain response, the susceptibility signal exhibits characteristic features.

At low temperatures (\SI{200}{\kelvin}), the susceptibility remains approximately constant upon loading and unloading, confirming a predominantly martensitic state. As the starting temperature increases (and thus the austenitic phase fraction increases), the magnitude of the susceptibility change under increasing uniaxial load grows, reaching its maximum conversion at \SI{260}{\kelvin}, where the austenitic phase fraction within the hysteresis is largest. Upon unloading, the susceptibility remains constant because no reverse transformation from martensite back to austenite occurs.
    
At \SI{270}{\kelvin}, the susceptibility increase only begins above about \SI{200}{\mega\pascal}, aligning with the expectation that higher stresses are required to initiate conversion at temperatures above the zero-stress hysteresis.
    
Upon unloading, a partial reverse transformation, indicated by a decrease in susceptibility, is observed starting at approximately \SI{50}{\mega\pascal} for temperatures between 260 and \SI{280}{\kelvin}. Above \SI{280}{\kelvin}, a transformation-induced hysteresis is no longer visible.
    
To estimate the degree of stress-induced transformation, we use the difference in susceptibility between the cooling and heating curves under \SI{20}{\mega\pascal} [Fig.~\ref{fig:PIT_Stress_strain(T)}(b)] as a reference for complete thermal transformation. By comparing the measured susceptibility change after the \SI{250}{\mega\pascal} load cycle at different temperatures to this reference, we calculated the percentage of stress-induced martensitic conversion.

Figure \ref{fig:PIT_Stress_strain(F)}(c) shows the result of this calculation. A maximum load of 250MPa is not sufficient to fully transform the PIT sample, achieving only a maximum conversion fraction of about \SI{43}{\percent} at 270K. For comparable bulk samples, 250MPa would be largely sufficient for complete conversion. This limited conversion is due to the load being predominantly absorbed by the rigid steel sheath. A simplified linear-elastic finite-element simulation of a PIT sample (radius 1.5mm, steel wall thickness 0.3mm) under 250MPa uniaxial load, performed without accounting for the magnetostructural transition, supports this. The stress transmitted to the core (about 100MPa) is only about one fifth of that carried by the sheath (about 500MPa) [Fig. \ref{fig:PIT_Stress_strain(F)}(d)], consistent with the applied load being predominantly carried by the stiffer steel sheath (200GPa) relative to the compliant porous Heusler core (60GPa). This finding underscores that achieving full transformation would require either an optimized composite geometry (e.g., thinner walls) or substituting the 316L steel with a more compliant sheath material

\subsubsection{Complex Strain Response}

Determining the phase conversion solely from the strain signal is problematic, as the strain behavior upon loading deviates significantly from expectations, suggesting a superposition with additional effects. This also suggests that the elastocaloric effect in similar composites cannot be reliably determined by simple strain measurements without further considerations. 

Comparing the PIT curves with the load cycle of the pure steel sheath at \SI{300}{\kelvin} [Fig.~\ref{fig:PIT_Stress_strain(F)}(a)], which shows linear, reversible elasticity and constant susceptibility, confirms that the linear elasticity of the sheath complicates the interpretation of the core's strain response. Despite this, the macroscopic strain changes correlate qualitatively with the susceptibility changes, though deviations in the exact onset pressures remain.

At \SI{200}{\kelvin}, a significant residual strain of about \SI{0.25}{\percent} remains after the loading cycle, even though the susceptibility suggests no structural transition. At higher temperatures (240 to \SI{270}{\kelvin}), residual strain is also visible after unloading, which aligns partially with the stress-induced austenite to martensite transition detected by the susceptibility. However, at \SI{280}{\kelvin}, where the susceptibility indicates a completely reversible response, the strain still exhibits residual deformation after the load cycle. A possible explanation is a small plastic deformation of the steel sheath, which would leave a permanent macroscopic strain without altering the magnetic state of the core. While the empty sheath deforms purely elastically at these loads [Fig.~\ref{fig:PIT_Stress_strain(F)}(a)], local stress concentrations at the core-sheath interface in the composite may promote limited plastic yielding not seen in the uniformly loaded empty sheath. Only at \SI{300}{\kelvin}, where the sample is fully austenitic, do both strain and susceptibility exhibit completely reversible behavior.

In addition to this mechanical contribution, a pressure-induced reorientation of the thermally induced martensite variants might contribute to this complex strain behavior. This reorientation changes the macroscopic strain but leaves the magnetic properties of the martensite largely unchanged, explaining the lack of correlation with susceptibility and the deviation from pure bulk behavior \cite{PFEUFFER2021117157}. Specifically, the pronounced non-linear strain change observed within the initial and final \SI{50}{\mega\pascal} ranges of the loading cycles could correspond to this reorientation process.

It is worth to consider that similar initial strain behavior can often be attributed to experimental artifacts, such as the gradual settling of non-parallel sample surfaces. All samples, including the steel reference, were prepared by the same grinding procedure, so that residual deviations from parallelism vary from sample to sample but remain comparable in magnitude. Nevertheless, this characteristic behavior was absent in the reference measurements of the pure steel sheath [Fig.~\ref{fig:PIT_Stress_strain(F)}(a)], which showed purely linear elastic behavior. This comparison suggests that geometric misalignment is unlikely to be the dominant cause.

Furthermore, this reorientation could also account for the absence of a clear reverse strain change upon heating [Fig.~\ref{fig:PIT_Stress_strain(T)}(a)], where the strain associated with austenite formation from the pressure-adapted martensite may be complexly superimposed and, therefore, not visible. Further microscopic investigations, such as in situ X-ray diffraction under load, are required to confirm this mechanism.

\section{\label{sec:Conclusion}Conclusion}

In this work, we demonstrated that mechanical confinement via powder-in-tube processing provides an effective strategy to maintain the integrity of the functional material, as the sheath confines even the fractured elastocaloric Heusler alloy in place, despite its intrinsic brittleness. The resulting composite enables reliable operation under high cyclic loads while preventing catastrophic failure.

The observed mechanical robustness includes stable operation over \num{100000} compressive load cycles at \SI{250}{\mega\pascal}, with only localized damage at the sample ends and no loss of core-sheath integrity. Even
under overload conditions up to approximately \SI{700}{\mega\pascal}, the composite remains intact and deforms plastically rather than failing catastrophically, demonstrating the effectiveness of the confinement strategy.

By combining mechanical testing with simultaneous in-situ measurements of strain and AC magnetic susceptibility, we showed that the transformation behavior in these composites is governed by a heterogeneous internal stress field rather than by the macroscopic mechanical response alone. As a consequence, macroscopic strain and transformed phase fraction decouple, and the transformation cannot be deduced from mechanical data alone.

Quantitative analysis based on AC susceptibility data indicates that at \SI{250}{\mega\pascal} about \SI{43}{\percent} of the Heusler core undergoes stress-induced transformation. Finite-element simulations attribute this limited conversion to stress partitioning between sheath and core, resulting in significantly reduced effective stress within the functional material.

The rigid steel sheath thus absorbs most of the applied load, limiting the achievable transformation fraction and identifying composite geometry and sheath material as the key levers for further optimization.

Overall, powder-in-tube processing provides a viable route toward fatigue-resistant caloric composites, and the combined in-situ approach provides quantitative access to the transformation behavior under realistic loading conditions. A direct calorimetric characterization of the confined core, in particular the adiabatic temperature change under load, is the natural next step toward assessing the caloric performance of these composites. Mechanical confinement thereby opens a broader design space for fatigue-resistant solid-state cooling.

\section*{Declaration of competing interest}

The co-author Oliver Gutfleisch is an Editor for Acta Materialia and was not involved in the editorial review or the decision to publish this article.
\newline

\begin{acknowledgments}

This work was supported by the Clean Hydrogen Partnership and its members within the framework of the project HyLICAL (Grant No. 101101461) as well as by the Deutsche Forschungsgemeinschaft (DFG) through the Würzburg-Dresden Cluster of Excellence on Complexity, Topology and Dynamics in Quantum Matter -- ctd.qmat (EXC 2147, Project No. 390858490), BEsT (Project-ID 456263705), and the CRC/TRR 270 (Project-ID 405553726). We acknowledge support by the HLD at HZDR, member of the European Magnetic Field Laboratory (EMFL).

\end{acknowledgments}

\bibliographystyle{elsarticle-num}
\bibliography{literatur}%

\end{document}